\documentclass[12pt]{iopart}
\usepackage{iopams}  
\usepackage{graphicx}
\usepackage{url}
\usepackage{harvard}
\citationmode{abbr}
\usepackage{xcolor}

\begin{document}
\renewcommand{\harvardurl}{URL: \url}
\newcommand{\citet}{\citeasnoun}
\newcommand{\netname}{PPNN\ }
\newcommand{\netnamenospace}{PPNN}

\title{Proton path reconstruction for pCT using Neural Networks}

\author{T. Ackernley$^{1,2}$, G. Casse$^{1,2}$ and M. Cristoforetti$^2$}

\address{$^1$Oliver Lodge, Department of Physics, University of Liverpool, Oxford Street, L69 7ZE Liverpool, United Kingdom \\
$^2$Fondazione Bruno Kessler (FBK), Via Sommarive, 18, Povo, 38123, Trento, Italy}
\ead{mcristofo@fbk.eu}
\vspace{10pt}
\begin{indented}
\item[]September 2020
\end{indented}

\begin{abstract}
The Most Likely Path formalism (MLP) is widely established as the most statistically precise method for proton path reconstruction in proton computed tomography (pCT). However, while this method accounts for small-angle Multiple Coulomb Scattering (MCS) and energy loss, inelastic nuclear interactions play an influential role in a significant number of proton paths. By applying cuts based on energy and direction, tracks influenced by nuclear interactions are largely discarded from the MLP analysis. In this work we propose a new method to estimate the proton paths based on a Deep Neural Network (DNN). Through this approach, estimates of proton paths equivalent to MLP predictions have been achieved in the case where only MCS occurs, together with an increased accuracy when nuclear interactions are present. Moreover, our tests indicate that the DNN algorithm can be considerably faster than the MLP algorithm.
\end{abstract}

%
%
%
%
%

\section{Introduction}



When reviewing recent developments in cancer treatment, proton beam therapy has seen rapid growth as an external beam radiotherapy technique, being increasingly favoured over traditional x-ray treatment for several tumours.
Unlike in regular radiation treatment, protons deposit most energy near the end of their path, a well-established effect known as the Bragg peak. By exploiting this property, protons are used to target tumours while subjecting their surroundings to little or no damage. Such treatment is well suited for tumours located near sensitive organs or in young patients for whom excess radiation exposure is a significant long term concern (\citet{tian2018evolution}, \citet{hu2018proton}, \citet{foote2012clinical}). Its capacity for depositing large amount of energy in a small volume increases the precision of treatment but so too the need to precisely locate the proton beam spot.

Accurate calibration of proton ranges relies on a detailed knowledge of the Relative Stopping Power, or RSP, of any tissue a proton will pass through along its path. Inaccurate placement of Bragg peaks can not only result in under-dosage of the target but also in significant exposure to the sensitive areas whose presence warranted proton therapy initially. Satisfactory resolution of RSP remains a substantial obstacle in unlocking the full potential of proton therapy. Current treatment planning systems rely on converting x-ray linear attenuation coefficient measurements, made in Hounsfield Units (HU), to RSP. Unfortunately, the non-unique relationship between HU and RSP introduces errors in the range of $~2-5$\%~\cite{beaton2019rapid}.

Proton computed tomography, or pCT, has been suggested as an alternative to overcome this problem. For proton therapy planning pCT offers the advantage of measuring proton RSP directly, removing conversion uncertainties by using the same particle for both planning and treatment~(\citet{doolan2015patient}).

For a given proton $i$, the line integral of the RSP is related to the energy loss using
\begin{equation*}
  \textrm{WEPL}_i \equiv \int_{\Gamma_i}\textrm{RSP}(x)\textrm{d}x\approx \int_{E_i^{out}}^{E_i^{in}}\frac{\textrm{d}E}{S_{water}(E)}
\end{equation*}
where $\Gamma_i\subset \mathbb{R}^3$ is the proton path, RSP(x) is the stopping power relative to water at position $x\in R^3$, $E^{in}_i$ and $E^{out}_i$ are the entrance and exit proton energies, and $S_{water}(E)$ is the stopping power of water for energy $E$. This integral is the Water Equivalent Path Length (WEPL). Starting from this equation, the pCT reconstruction problem can be mapped to that of reconstructing each individual protons path, combined with the calculation of WEPL (through the right side of the equation), to recover the RSP map. It is therefore crucial that the reconstruction of the proton path will be as accurate as possible. Indeed, the better the determination of the proton trajectories, the better the RSP calculation will be. 

Image reconstruction using protons poses an additional challenge over standard x-ray CT: during passage through matter protons experience significant deflections through Multiple Coulomb Scattering (MCS), and, more rarely, nuclear interactions, resulting in non-trivial curved paths. The probability of nuclear reactions compared to ionization interactions is less than $1\%$ for 200 MeV protons. As a consequence, the influence of nuclear interactions of
protons with atomic nuclei can be treated as correction to the
electromagnetic processes~(\citet{fippel2004monte}). Accurate reconstruction of these paths determines the achievable imaging resolution in proton computed tomography (pCT) and thus the exact dose distribution in proton therapy. Unlike with x-ray CT, in which photon number attenuation along straight propagation lines is considered, the pCT reconstruction process requires proton paths to be individually estimated to account for the curved trajectories if an improved resolution is to be achieved~(\citet{johnson2017review}).

This requirement excludes direct reuse of many well-developed image reconstruction methods developed in x-ray CT~(\citet{johnson2017review}, \citet{bovik2009essential}). Iterative algebraic methods, such as the algebraic reconstruction technique (ART), have been proposed as plausible pCT image reconstruction methods~(\citet{li2006reconstruction}, \citet{johnson2017review}), but the computational cost of these algorithms is considerably high. More efficient techniques are direct reconstruction methods, often following on from x-ray CT methods, who's development is an active area of research, as discussed in \citet{khellaf2020comparison}.

At the core of these methods is the Most Likely Path (MLP) formalism for the reconstruction of the single proton trajectory. While scattering remains an inherently probabilistic process, precluding the exact prediction of any single track, MLP is well established as the most statistically precise method to account for MCS processes~(\citet{schulte2008maximum}, \citet{williams2004most}, \citet{fekete2015developing}). Since its introduction in 1994~(\citet{schneider1994multiple}), the MLP formalism as presented in \citet{schulte2008maximum} has undergone various refinements for use in different application scenarios (\citet{fekete2015developing}, \citet{collins2017theoretical}, \citet{collins2017extension}, \citet{krah2019polynomial}, \citet{brooke2020inhomogeneous}).

In addition to the entry and exit positions of the beam, the MLP algorithm utilises the angle between the direction of travel and the perpendicular to the phantom surface to significantly improve the prediction~(\citet{schneider1994multiple}). These quantities can be measured by modern pCT scanners systems~(\citet{johnson2017review}). However, while the formulation of MLP accounts for small-angle multiple Coulomb scattering (MCS) and small energy loss, nuclear interactions play an influential role in a significant number of proton trajectories~(\citet{johnson2017review}).
Recommended practice is therefore to reduce the events influenced by nuclear interactions or large angle MCS through a $3\sigma$ cut on both the difference in energy and the difference in the direction of travel angle between entry and exit~(\citet{schulte2008maximum}). Unfortunately, this results in a reduction of the protons available for the pCT image reconstruction and in an increase of the time needed to compute the relative stopping power map for proton therapy treatment planning. The need to estimate proton paths on a one by one basis, coupled with the inability to use many well-established x-ray CT reconstruction methods, comes with a significant computational burden~(\citet{johnson2017review}). Various avenues of research into overcoming this problem have been explored, from optimizing the computer code for MLP evaluation~(\citet{mcallister2009general}), to alternative approaches approximating MLP through cubic splines~(\citet{fekete2015developing}) or polynomial approximations~(\citet{krah2019polynomial}).

It is in this context that we introduce a new and original approach for the estimation of the proton paths based on Machine Learning, through utilisation of a Deep Neural Network. The Proton Path Neural Network (\netnamenospace) is capable of reaching the same performance as MLP when this last is applicable, and exceeding it on a large fraction of paths influenced by nuclear interactions. Moreover, our tests indicate that \netname exhibits significantly shorter execution time than the MLP approach.

The paper is organised as it follows. An overview of the Monte Carlo simulations used and the relevant physics environment is given in Section~\ref{sec:mc_simul}. This is followed in Section~\ref{sec:mlp} by a description of the existing MLP proton path reconstruction, before the introduction of \netname in Section~\ref{sec:network}. Studies comparing the reconstruction capabilities of \netname against MLP are presented in Section~\ref{sec:rmse}, with further analysis into the methods' behavioural differences and the characteristics of corresponding tracks introduced in Section~\ref{sec:error_func_of_deviations} and  Section~\ref{sec:diff_traj_for_diff_errors} respectively. Initial work investigating performance on an inhomogeneous phantom is reviewed in Section~\ref{sec:inhomo}. Comparison of execution times is covered in Section~\ref{sec:times}. Finally, a discussion of these results is presented in Section~\ref{sec:discussion}.

\section{Materials and methods}

\subsection{Monte Carlo Simulation}\label{sec:mc_simul}

The Monte Carlo simulations presented were performed using GATE v9.0 (\citet{jan2011gate}), a framework built upon the widely used Geant4 10.6 Monte Carlo simulation toolkit~(\citet{agostinelli2003geant4}). Simulations incorporating only electromagnetic processes were performed using the $emstandard$ physics list. The impact of nuclear interactions, among a full regime of physics processes, were modeled using the $QGSP\_BIC$ physics list. In the discussion of the results, the choice of physics environment is indicated for each simulation.

\begin{figure}[t]
\centering
\includegraphics[width=1\linewidth]{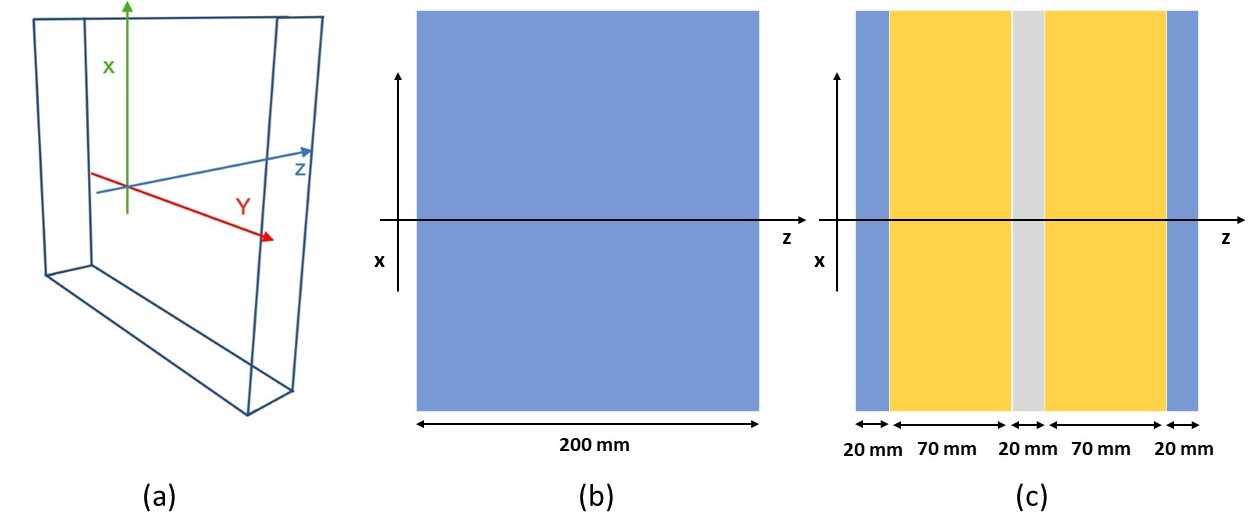}
	\caption{Illustration of the Monte Carlo geometry used in this study. 3D representation of the phantom space (a) and 2D projection on the $x-z$ plane for the water (b) and inhomogeneous phantom (c). Trajectories are only scored and monitored within the phantom volume itself. Note that for convenience we redefine our coordinate axis such that the initial point of each trajectory is located at the origin.}
	\label{fig:mc_setup}
\end{figure}

Our main model consists of a sheet of water centred on the origin of a standard x-y-z coordinate system with a side length of $20\,$cm in the z-axis direction and arbitrarily large extents in x and y. Monoenergetic protons initialised at $200$MeV are simulated through the phantom, originating at the central point of the phantom's $z= -10\,$cm face, such that their initial direction of travel are orientated inwards and perpendicular to the face and parallel to the positive z-axis direction. For convenience in the following we redefine our coordinate axis such that the initial point of any trajectory is located at the origin, with particles initialised at a depth of $0\,$cm and extending in range to a depth of $20\,$cm. This arrangement is illustrated in Figure \ref{fig:mc_setup}. 

Each data set produced initially contained $10^{6}$ events; however, only trajectories which traversed the full phantom depth were retained, reducing the number of events ultimately used. Typically this led to data sets in excess of $800,000$ events. For the purposes of this study, trajectories themselves are quantified as a series of spatial coordinates evenly distributed at $0.1\,$cm intervals, including both phantom faces. A total of $201$ coordinate points represent a complete path through the phantom, consisting of $603$ variables. As the $z$ depth coordinates are therefore a fixed set of values shared by all trajectories, for predicting a track only the $x$ and $y$ variables need be considered. Similarly, the initial and final points of each trajectory are known for each track and so likewise neglected. Thus a track prediction consists of two sets of $199$ points each, for a total of $398$ variables per track.

In addition, as a first check of the robustness of the \netname approach in inhomogeneous media, the procedure as stated was repeated using a phantom comprising $2$~cm of water, $7$~cm of skull, $2$~cm of cortical-bone, $7$~cm of skull, and $2$~cm of water. For the purposes of this simulation, cortical-bone was defined using material data found in \citet{icru1993report49}. Due to the increased stopping power, to ensure that a large fraction of impinging protons successfully traverse the phantom's full length, a beam energy of $230$~MeV was used. Additional simulations with an equivalently sized water phantom were carried out as before at this initial proton energy, as a baseline for comparison. All $230$~MeV simulations were carried out under the $QGSP\_BIC$ physics list.

\subsection{Most Likely Path}\label{sec:mlp}

Given the coordinate system and the simulation framework described in Section~\ref{sec:mc_simul}, with the proton beam directed along the $z$ direction, at any given depth along $z$ a proton's path can be characterised by the two coordinates $x$ and $y$ and the two angles $\theta$ and $\phi$ relative to the $z$-axis. Proton scattering can be considered independent along the $x$ and $y$ axis and the MLP can be expressed independently for the two 2D parameter vectors $\mathbf{x}=(x,\theta)$ and $\mathbf{y}=(y,\phi)$.

Considering $\mathbf{x}$ for example, from \citet{schulte2008maximum} the MLP of protons in a homogeneous medium can be expressed, in a Gaussian approximation of the generalised Fermi-Eyeges theory of Multiple Coulomb Scattering (MCS), as
\begin{equation}
	\mathbf{x}_{\textsc{mlp}}(z)=(\Sigma_1^{-1}+R_1^T\Sigma_2^{-1}R_1)^{-1}(\Sigma_1^{-1}R_0\,\mathbf{x}_{in}+R_1^T\Sigma_2^{-1}\,\mathbf{x}_{out}),
\end{equation}
where $\mathbf{x}_{in}$ and $\mathbf{x}_{out}$ are the relevant entry and exit coordinates in the two 2D parameter vectors as mentioned above, $R_0$ and $R_1$ are the change of basis for small-angle rotation matrices
\begin{eqnarray}
	R_0 = \left(\begin{array}{cc}
  1 & z-z_{in} \\
  0 & 1 
\end{array}\right) &,\ \ & 	
R_1 = \left(\begin{array}{cc}
  1 & z_{out}-z \\
  0 & 1 
\end{array}\right), 
\end{eqnarray} 
and $\Sigma_1$ and $\Sigma_2$ are covariance matrices
\begin{eqnarray}
	\Sigma_1 = \left(\begin{array}{cc}
  \sigma^2_{t_1} & \sigma^2_{t_1\theta_1} \\
  \sigma^2_{t_1\theta_1} & \sigma^2_{\theta_1} 
\end{array}\right) &,\ \ & 	
\Sigma_2 = \left(\begin{array}{cc}
  \sigma^2_{t_2} & \sigma^2_{t_2\theta_2} \\
  \sigma^2_{t_2\theta_2} & \sigma^2_{\theta_2} 
\end{array}\right), 
\end{eqnarray} 
with components, called \textit{scattering moments}, given for $\Sigma_1$ by the integrals
\begin{eqnarray}
	\sigma^2_{t_1} = E_0^2\left(1+0.038\ln \frac{z-z_{in}}{X_0}\right)^2\int_{z_{in}}^z \frac{(z - u)^2}{\beta^2(u)p^2(u)}\frac{\textrm{d}u}{X_0}\\
	\sigma^2_{\theta_1} = E_0^2\left(1+0.038\ln \frac{z-z_{in}}{X_0}\right)^2\int_{z_{in}}^z \frac{1}{\beta^2(u)p^2(u)}\frac{\textrm{d}u}{X_0}\\
	\sigma^2_{t_1\theta_1} = E_0^2\left(1+0.038\ln \frac{z-z_{in}}{X_0}\right)^2\int_{z_{in}}^z \frac{(z - u)}{\beta^2(u)p^2(u)}\frac{\textrm{d}u}{X_0},
\end{eqnarray}
where $u$ is the predicted proton path. The equivalent scattering moments for $\Sigma_2$ are found by replacing $z_{in}$ with $z$ and $z$ with $z_{out}$ in the equations above. $\mathbf{y}_{\textsc{mlp}}(z)$ follows identically, with $\mathbf{x}_{in}$ and $\mathbf{x}_{out}$ replaced by $\mathbf{y}_{in}$ and $\mathbf{y}_{out}$ as necessary.

Assuming a homogeneous phantom composed of water, we use $X_0=36.1\,$cm for the radiation length of the material and $E_0=13.6$~MeV. The momentum velocity ratio ${1}/{\beta^2(u)p^2(u)}$ is approximated with a fifth-order polynomial following \citet{schulte2008maximum}.
This quantity is specific to the proton energy used; implementation for other energies requires its recalculation for accurate performance. For protons at $230$~MeV this was calculated as outlined in \citet{schulte2008maximum}. Monoenergetic protons initially at the required energy were incident on a simulated $20$~cm deep water sample. The fifth-order polynomial was fitted to distribution of the mean value of ${1}/{\beta^2(u)p^2(u)}$ recorded at $5$~mm intervals throughout.  

\subsection{Proton Path Neural Network}\label{sec:network}

\begin{figure}[tbp]
  \centering
  \includegraphics[width=.8\linewidth]{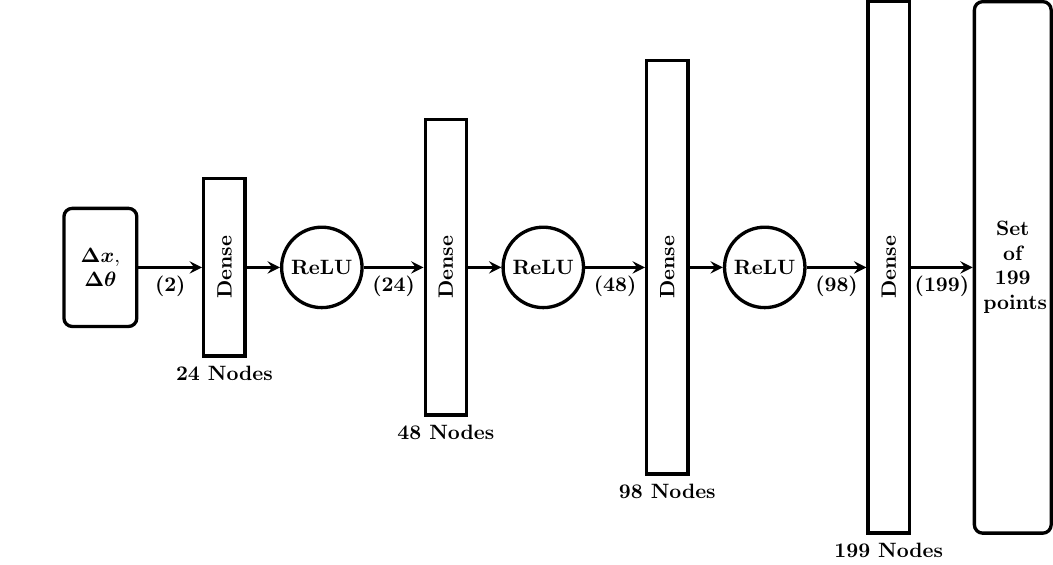}
    \caption{\netname architecture. The Proton Path Neural Network \netname consists of four fully connected layers with 24, 48, 96, 199 nodes and a Relu activation function after each of the first three layers. The current number of variables present at various points is additionally indicated in brackets.}
    \label{fig:PPNN}
  \end{figure}

The Proton Path Neural Network (\netnamenospace) is fully connected neural network based model designed to predict a proton trajectory in the form of a series of spacial points, as described in Section~\ref{sec:mc_simul}, using variables similar to those employed by MLP calculations. As with the MLP, trajectories along the $x$ and $y$ directions are reconstructed independently by separate instances of the same network. The input features of the network are quantities which can be recorded by a modern pCT scanning apparatus; $\Delta x = (x_{out} - x_{in})$ and $\Delta \theta = (\theta_{out} - \theta_{in})$ in the $x$ direction and equivalently $\Delta y = (y_{out} - y_{in})$, $\Delta \phi = (\phi_{out} - \phi_{in})$ along $y$. This data is passed through 4 fully connected (or dense) layers of 24, 48, 96 and 199 nodes respectively. This type of layers are the most simple between the many developed in the context of Deep Neural Network: 
the output of the layer is a vector $\mathbf{y}$ obtained by
\begin{equation*}
  \mathbf{y} = \sigma(\mathbf{W}\cdot\mathbf{x} + \mathbf{b})
\end{equation*}  
where $\mathbf{W}$ and $\mathbf{b}$ are called respectively weigths and bias and correspond to the parameters of the layer that will be fixed during training of the network;  $\mathbf{x}$ is the input vector and $\sigma$ is the activation function introducing non linear effects in the network behaviour.
As activation function we employed the Rectified linear unit (ReLU) after each of the first 3 layers ($\textrm{ReLU}\left( x\right) = \textrm{max}\left( 0,x \right)$).
A representation of the network architecture is presented in Figure~\ref{fig:PPNN}.

Training and validation of the network was performed using more than 1,600,000 trajectories (800,000 along each direction) generated as described in Section~\ref{sec:mc_simul} using the $QGSP\_BIC$ physics list. $80\%$ of the tracks are used for the training and the remaining $20\%$ reserved for validation. Optimization of the network weights is performed using the {\tt Adam} algorithm~(\citet{kingma2014adam}) with a learning rate fixed at $10^{-5}$. For the loss, the Mean Squared Error (MSE) is used,
\begin{equation}
    \textrm{MSE} = \frac{1}{M}\sum_m^M\frac{1}{N}\sum_n^N(u_{mn} - \hat{u}_{mn})^2,
\end{equation}
where $M$ is the number of samples, $N=199$ is the number of points in each proton path, $u$ again the predicted path and $\hat{u}$ the true trajectory. The (Square) Root of the Mean Squared Error (RMSE) is commonly adopted in literature evaluating the performance of the MLP reconstruction procedure. At a batch size of 32 samples per batch, one epoch (one cycle through the full training dataset) running on Tesla K80 GPU requires approximately 80 seconds on a Standard NC6 Microsoft Azure machine. For an introduction on Deep Neural Network we suggest looking at the free material available at https://d2l.ai/.

\begin{figure}[tbp]
  \centering
  \includegraphics[width=.7\linewidth]{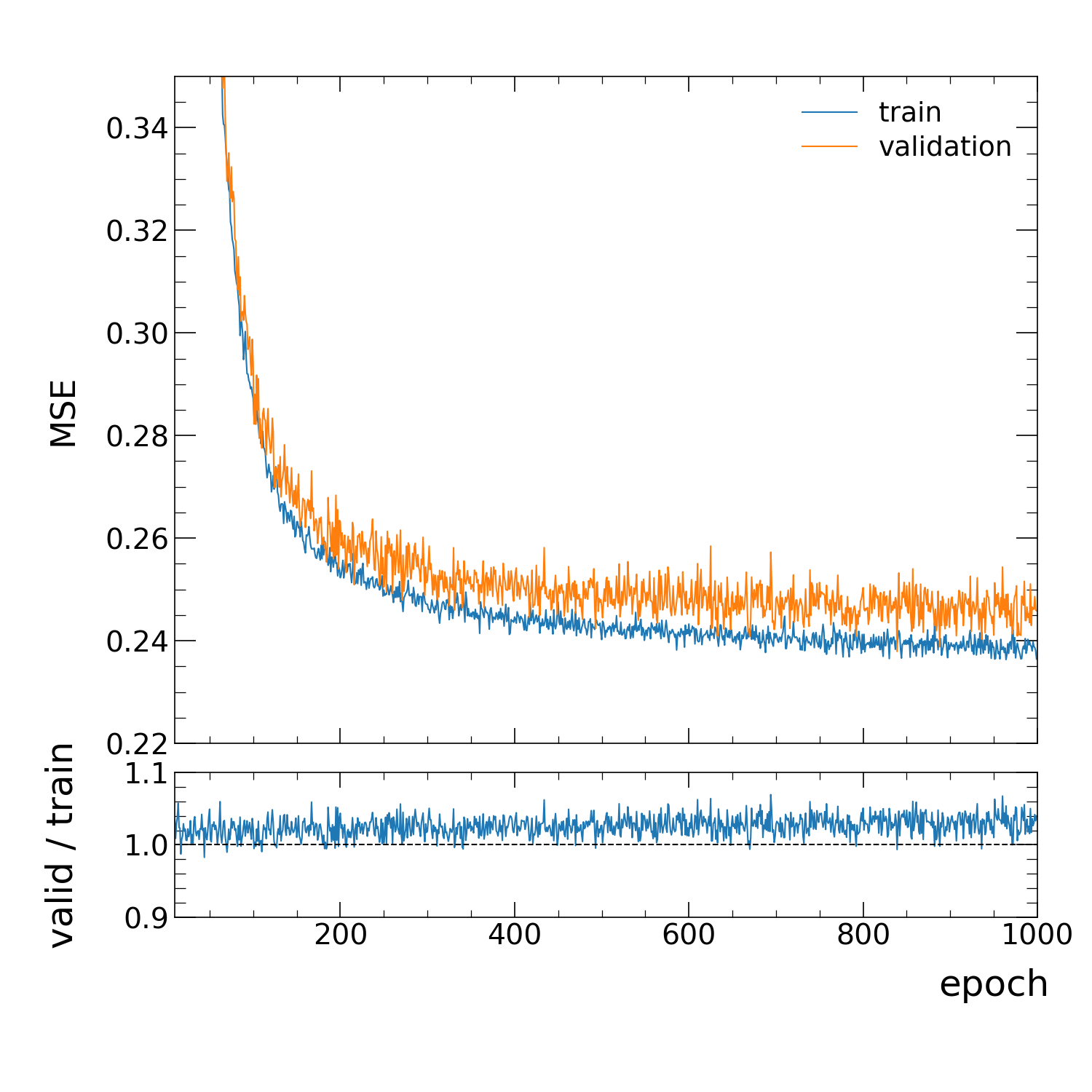}
    \caption{Loss history during network training at each epoch, for both the training and validation.}
    \label{fig:PPNN-hist}
  \end{figure}

The loss history can be seen in Figure~\ref{fig:PPNN-hist}, in which after around 400 epochs the loss flattens both for the train and validation datasets with the ratio between the two histories almost constant; suggesting that the network is not overfitting to the examples present in the training dataset. Ultimately the model was trained for 1000 epochs.

In addition, a second instance of the \netname was trained with a $230$~MeV proton dataset in excess of $1,400,000$ events, using the same methodology and a pure water phantom. This instance is used when reconstructing datasets with protons at that energy.

\section{Results}\label{sec:results}

To principally test the performance of \netname two entirely new datasets of 800,000 protons each were generated: the first with only electromagnetic interactions ($emstandard$ physics list), the other with all the physical processes including nuclear interactions ($QGSP\_BIC$ physics list). These data sets are generated independently from that used during the \netname training procedure to avoid any possible source of overfitting.

\subsection{Root Mean Squared Error}\label{sec:rmse}

\begin{figure}[t]
\centering
\begin{tabular}{cc}
\includegraphics[trim={0cm 2cm 0cm 0cm},clip,width=.5\linewidth]{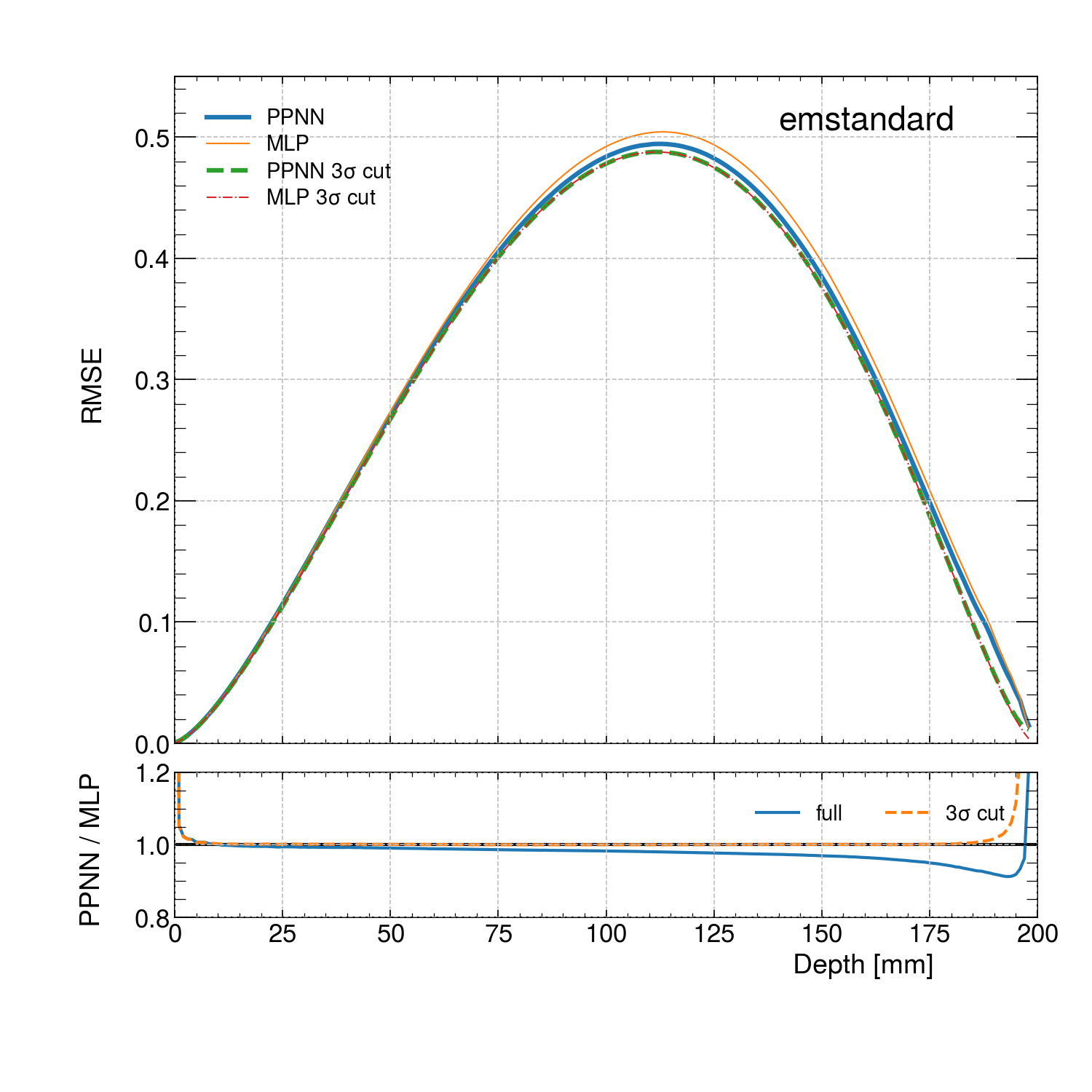}&
\includegraphics[trim={0cm 2cm 0cm 0cm},clip,width=.5\linewidth]{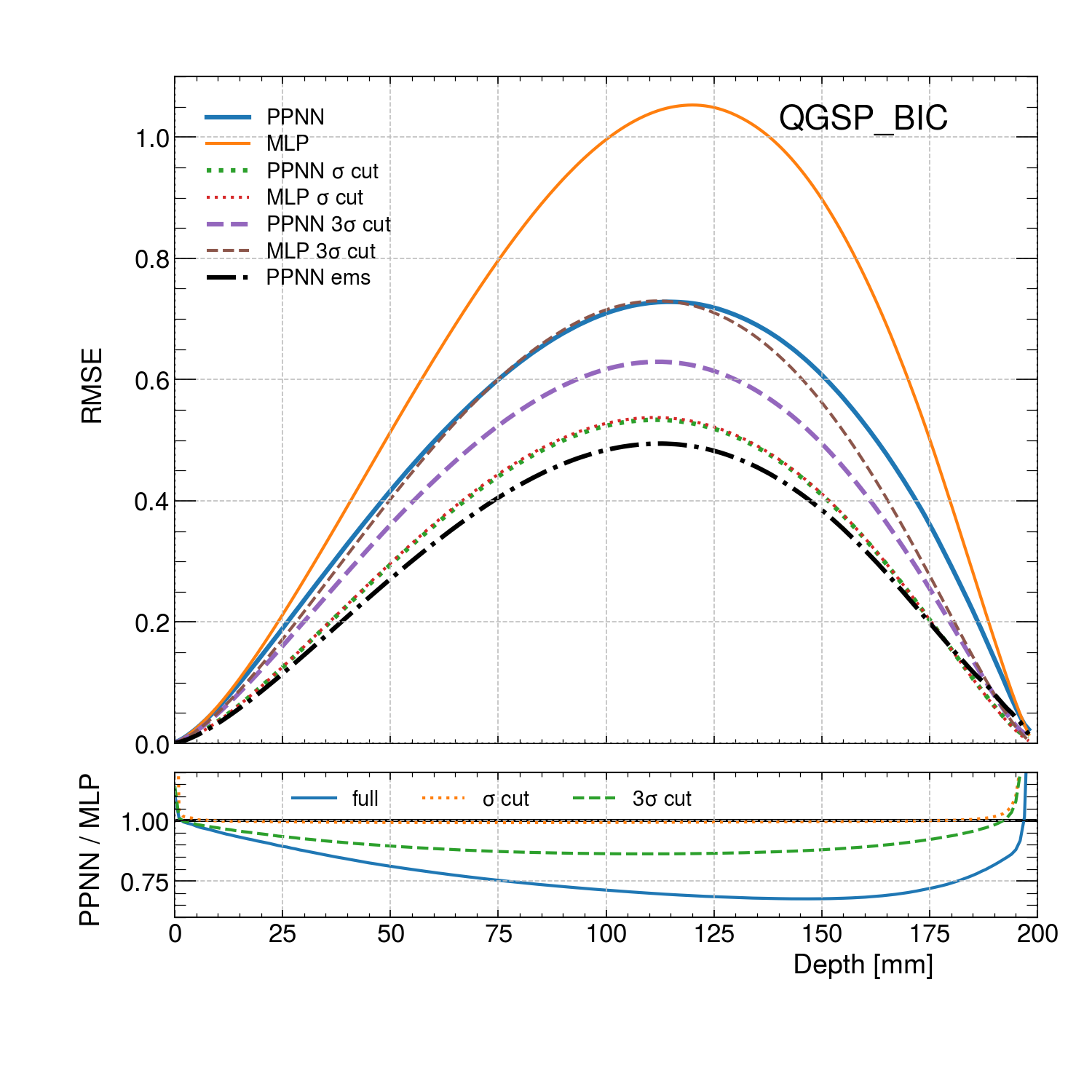}\\
(\textbf{a}) &
(\textbf{b})
\end{tabular} 
\caption{Root Mean Squared Error obtained with MLP and \netname using the (a) $emstandard$ and (b) $QGSP\_BIC$ datasets. Solid lines are the performance on the full dataset while dotted and dashed incorporate $1\sigma$, and $3\sigma$ cuts, performed on the energy and difference in the direction of travel angle between entering and exiting the phantom, respectively. The dashed-dotted line in (b) is the same solid \netname result in (a) added here to have a clear picture of the increasing of the errors when including nuclear interactions.} \label{fig:rmse}
\end{figure}

Figure~\ref{fig:rmse}-(a) shows the RMSE for estimates of the paths using \netname or MLP on the $emstandard$ dataset. Even without the $3\sigma$ cuts suggested in~\citet{schulte2008maximum} we can see that the difference between the two predictions is quite small. This difference disappears (the two lines corresponding to the MLP and \netname case are barely distinguishable) upon applying said $3\sigma$ cut to the angles and energy; under which here only $\sim 1\%$ of the paths are omitted. This result clearly shows that the \netname prediction is fully consistent with the MLP approach, indicating that the approximations inherent to the method are valid. This is crucial because anything different would represent a serious flaw in the \netname reconstruction method.

Moreover, the difference in the \netname prediction error with or without the cut is practically negligible, suggesting that our method can be applied to reconstruct trajectories where processes other than MCS are present. This is more evident in Figure~\ref{fig:rmse}-(b) where the RMSE is evaluated for the $QGSP\_BIC$ dataset. When nuclear interactions are included the error significantly increases, but to a far lesser extent for \netname than for MLP. Only with a $1\sigma$ cut do the performances of the two methods become comparable. Unfortunately, such a huge cut entails the loss of $\sim 24\%$ of the tracks. Comparing the full interaction dataset result with that of the pure electromagnetic result, we see that with the typical $3\sigma$ cut applied to both cases the RMSE of \netname is about $26\%$ larger for the full interaction that for the pure MCS dataset. For the $2\sigma$ cut the discrepancy in performance decreases to around $20\%$, which corresponds to a fraction of discarded tracks of $\sim 8\%$ from the $QGSP\_BIC$ dataset.

\subsection{Error as a function of deviations}\label{sec:error_func_of_deviations}

\begin{figure}[t]
\centering
\begin{tabular}{cc}
  \includegraphics[trim={0cm 1.5cm 0cm 0cm},clip,width=.5\linewidth]{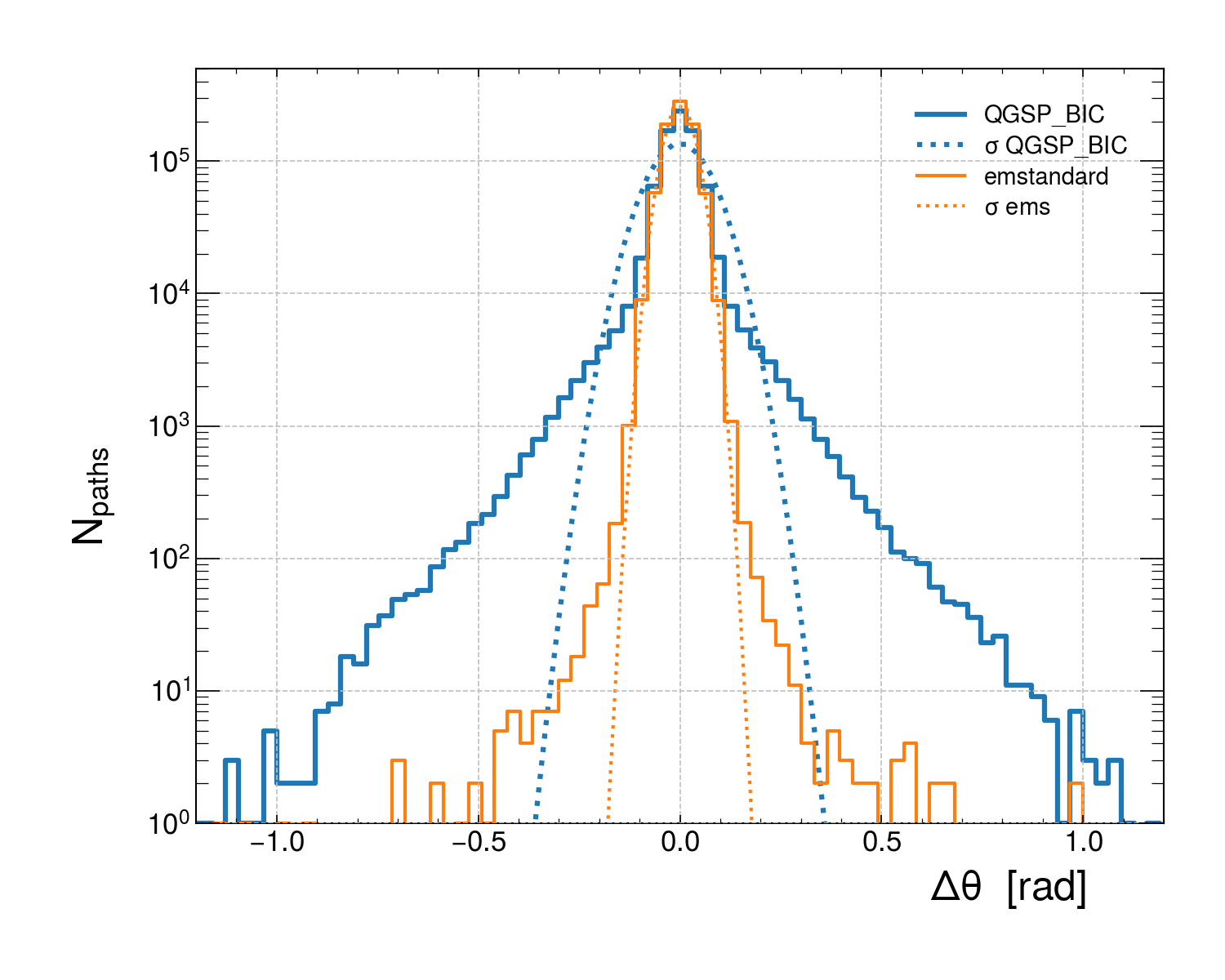}&
\includegraphics[trim={0cm 1.5cm 0cm 0cm},clip,width=.5\linewidth]{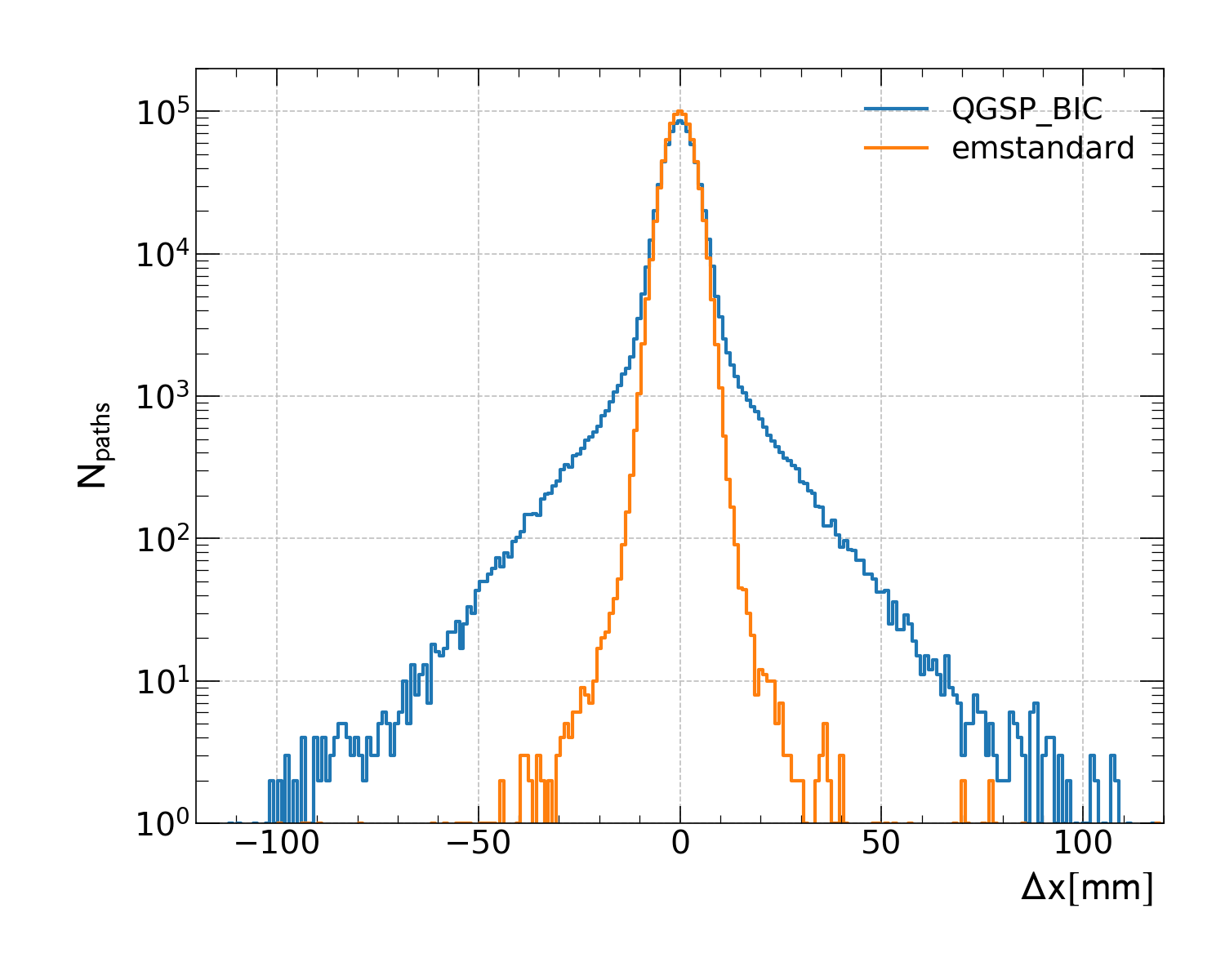}\\
(\textbf{a})  &
(\textbf{b}) 
\end{tabular}
\caption{(a) Distribution of  $\Delta\theta = (\theta_{out}-\theta_{in})$ angle for the two test datasets (solid lines) overlaid with the associated Gaussian using the $\sigma$ values obtained from a fit of the $emstandard$ data and the $QSPG\_BIC$ data (dotted lines). (b) Distribution of $\Delta x =(x_{out} - x_{in})$. In both plots it is evident that an exponential rather than a Gaussian decay provides a better fit with respect to the number of paths for the $QSPG\_BIC$ dataset.} \label{fig:delta}
\end{figure}

\begin{figure}[htb]
\centering
\begin{tabular}{c}
\includegraphics[width=.7\linewidth]{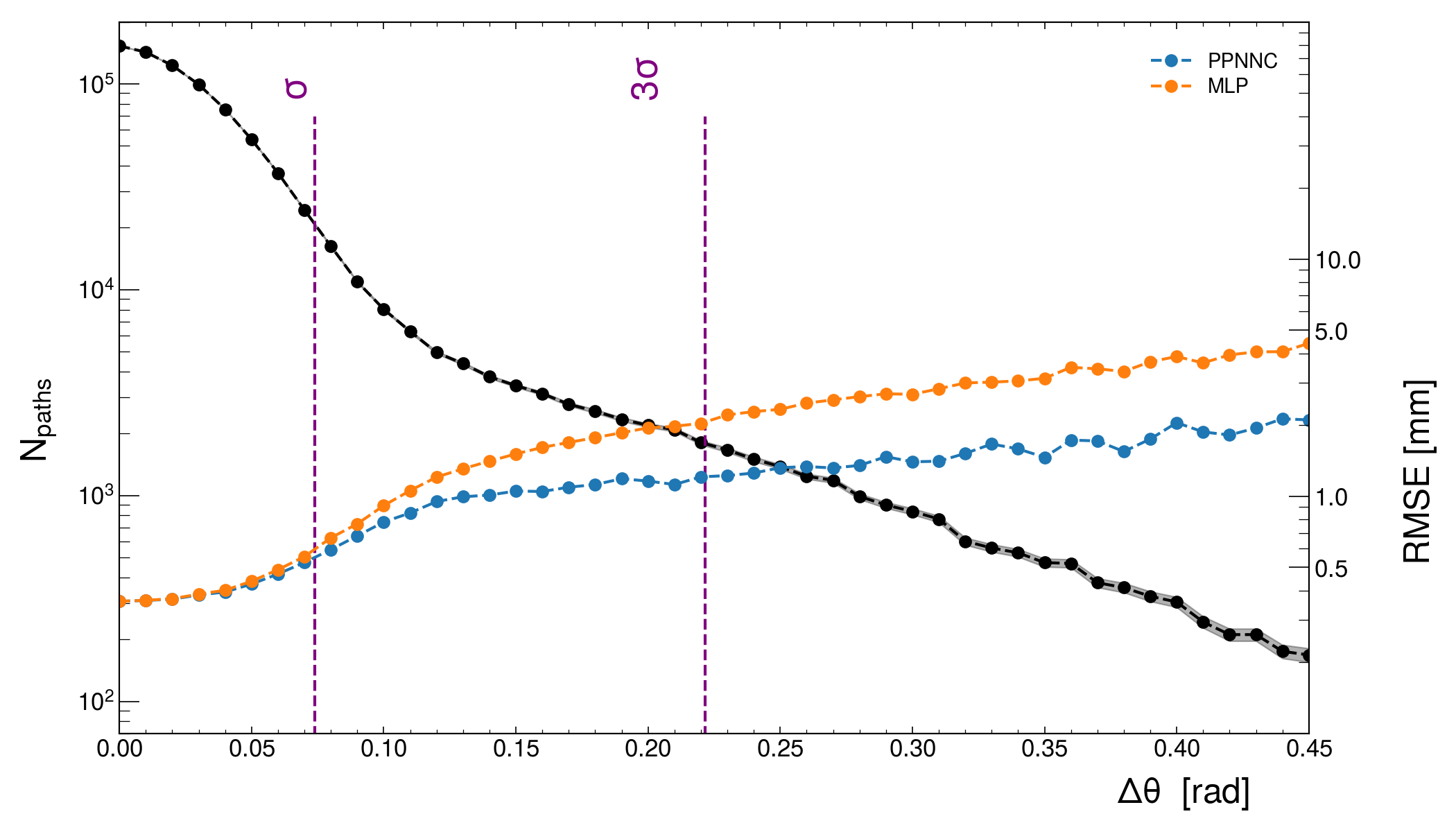}\\
(\textbf{a})  \\
\includegraphics[width=.7\linewidth]{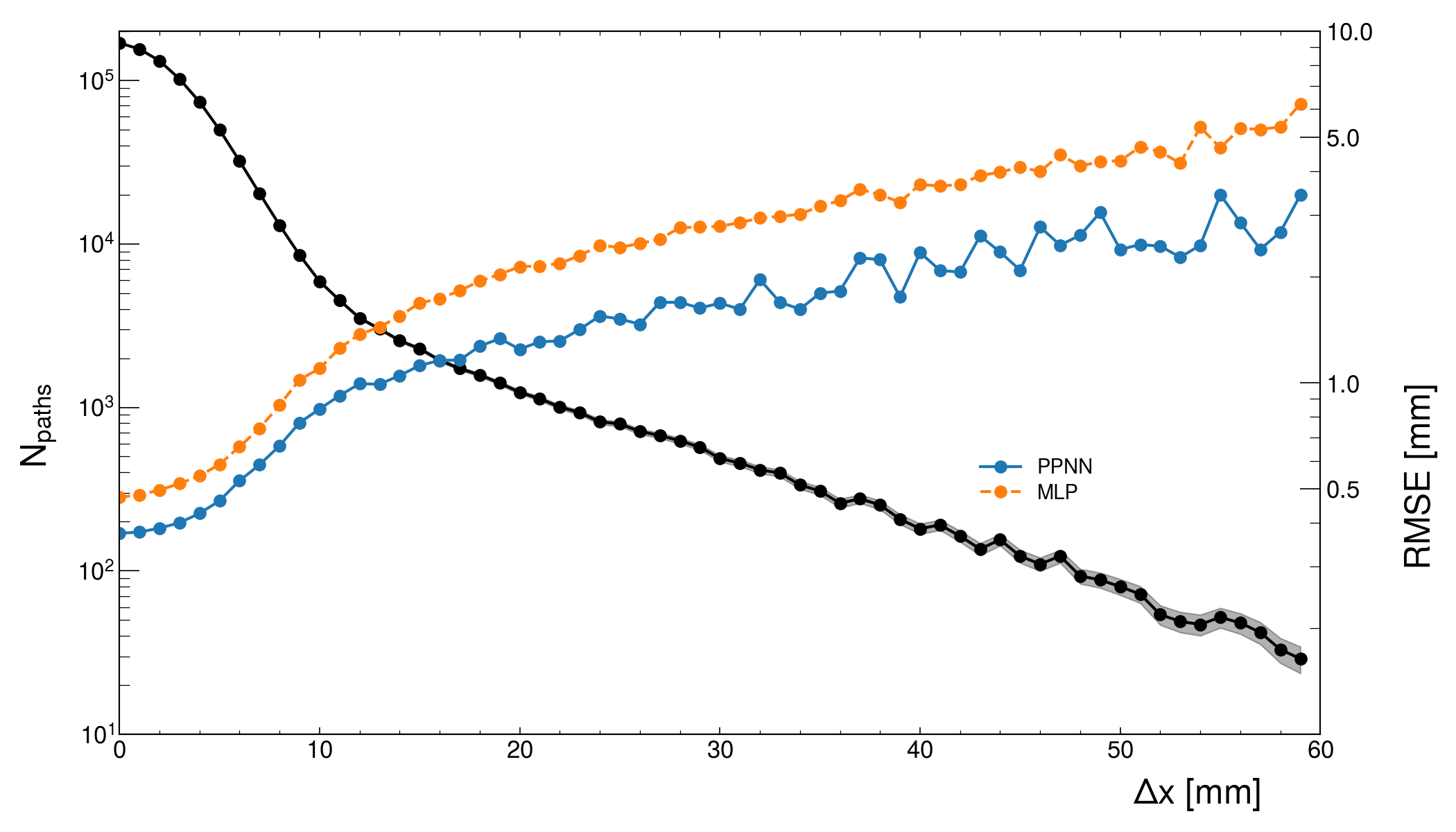}\\
(\textbf{b}) 
\end{tabular}
\caption{(a) RMSE (right vertical axis, coloured lines) and number of paths (left vertical axis, black lines) as a function of $\Delta\theta$ for \netname and MLP evaluated on the $QSPG\_BIC$ dataset. The shaded black area represent the statistical error. Vertical lines refer to the position of the 1 and 3 $\sigma$ cut. (b) Same as (a) but as a function of $\Delta x$. The difference in performance between the two methods emerges immediately.} \label{fig:delta_error}
\end{figure}

To understand the origin of this difference in performance between the two methods, Figure~\ref{fig:delta}-(a) illustrates the distribution of $\Delta\theta = (\theta_{out} - \theta_{in})$ for both $QSPG\_BIC$ and $emsstandard$ datasets. The $\sigma$ cut is applied assuming a Gaussian distribution of the signal, but from the figure a difference between the two distributions clearly emerges. For the full physics simulation the Gaussian approximation, as employed in the MLP, clearly fails to describe the distribution. While the cuts based on a Gaussian fit are acceptable in the $emstandard$ case, they exhibit a large discrepancy with data when the full range of physics processes are included. In Figure~\ref{fig:delta}-(b) we see a similar result for the distribution of lateral displacement $\Delta x = (x_{out}-x_{in})$, with the Gaussian shape of the $emstandard$ distribution supplanted by an exponential decrease in the $QSPG\_BIC$ distribution.

Given this observation and having verified that the \netname approach has the same perfomances as MLP in the context of pure electromagnetic interaction, where MLP is designed to work, from now on we will consider only the results obtained using the $QSPG\_BIC$ physics dataset as a much more realistic representation of clinical pCT scenario. 

As the distributions of Figure~\ref{fig:delta} clearly show the limits of the MLP formulation, it is interesting therefore to consider how the error increases as a function of the two variables $\Delta\theta$ and $\Delta x$. This is presented in Figure~\ref{fig:delta_error}.
Here the proton paths are collected into bins of $0.1$ rad and $1$ mm for $\Delta \theta$ and $\Delta x$ respectively, with the RMSE computed in the corresponding direction. The figure compares the error (right axis) and the number of trajectories (left axis) to show the differences in performance. Note the logarithmic scale on both right and left $y$ axis. From Figure~\ref{fig:delta_error}-(a) we see that, as expected from the RMSE plot, the two lines for \netname and MLP begin to separate at around $1\sigma$ cut at $\Delta \theta\simeq 0.075$ rad. For $35\%$ of the tracks $\Delta \theta$ is larger than 0.075, implying that the \netname method improves on the MLP reconstruction for an important fraction of proton paths. Notice that the same analysis must be done for the $\phi$ angle which would remove an analogous number of paths, resulting in a final cut of almost $50\%$ of the tracks. Figure~\ref{fig:delta_error}-(b) shows the reconstructed paths distribution broken down in term of final displacement, $\Delta x$. Again the performance of \netname  is consistently better across the full span of the plot, with trajectories at large angle deviations resolved with improved precision.

\subsection{Different trajectories for different errors}\label{sec:diff_traj_for_diff_errors}

\begin{figure}[t]
\centering
\includegraphics[trim={0cm 1.5cm 0cm 0cm},clip,width=.5\linewidth]{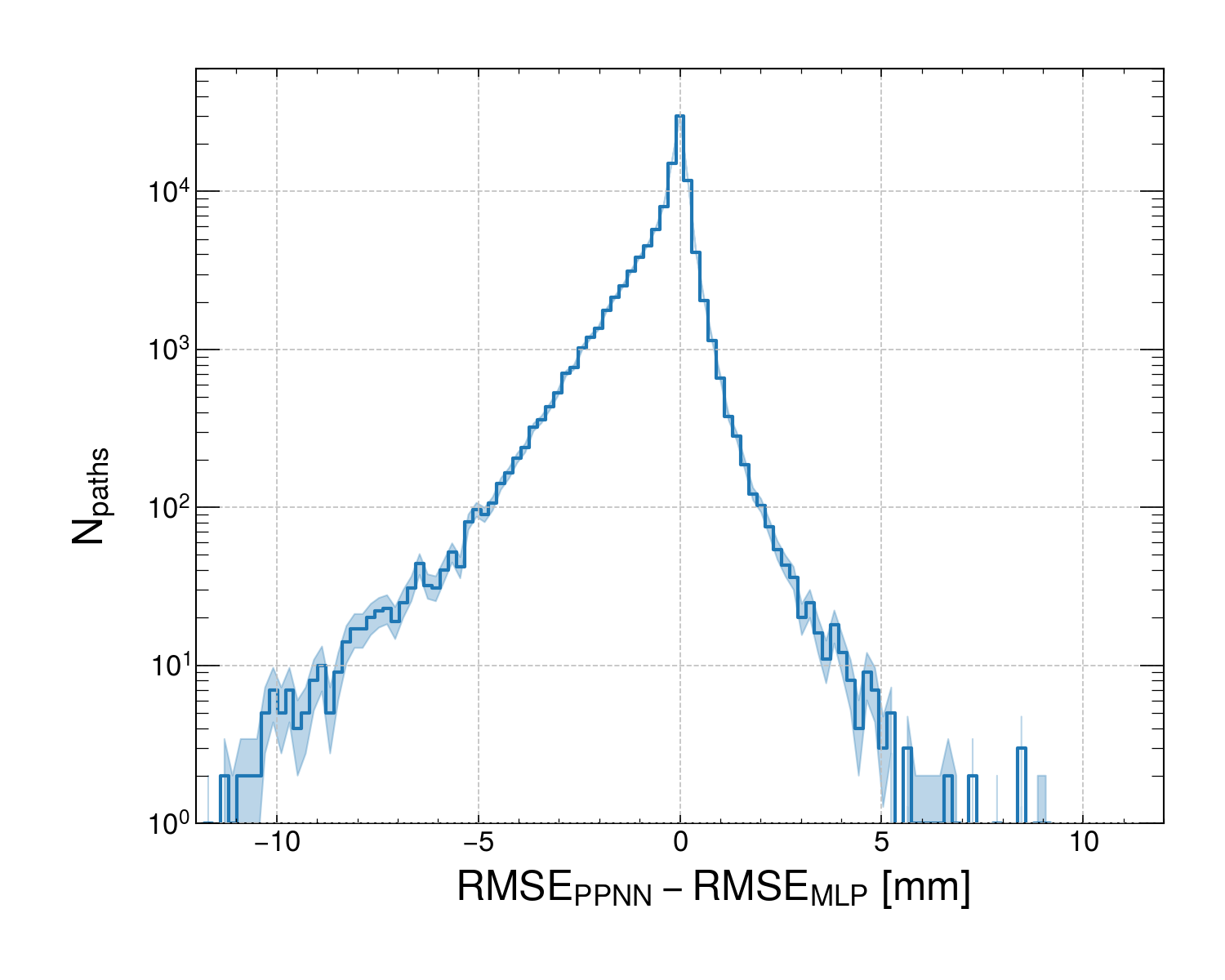}
\caption{Distribution of the difference between the RMSE of \netname and MLP for the $QSPG\_BIC$ dataset. The shaded area correspond to the statistical error.} \label{fig:delta_RMSE}
\end{figure}

To gain an insight into the tracks with the largest difference in reconstruction performance, let us begin by considering only tracks outside the $1\sigma$ cut in $\theta$. In Figure~\ref{fig:delta_RMSE} we present the distributions of the difference between the RMSE for \netname and MLP for tracks outside the aforementioned cut. Negative values of the difference correspond to tracks in which \netname had the smallest error, while the positive side of the axis corresponds to the inverse. In the first instance we can see that the profile is exponential, while in the second the decay is noticeably faster; confirming that at large deviations of the angle $\theta$, \netname shows a notably superior performance.

\begin{figure}[t]
\centering
\begin{tabular}{cc}
\includegraphics[trim={0cm 2.5cm 0cm 0cm},clip,width=.5\linewidth]{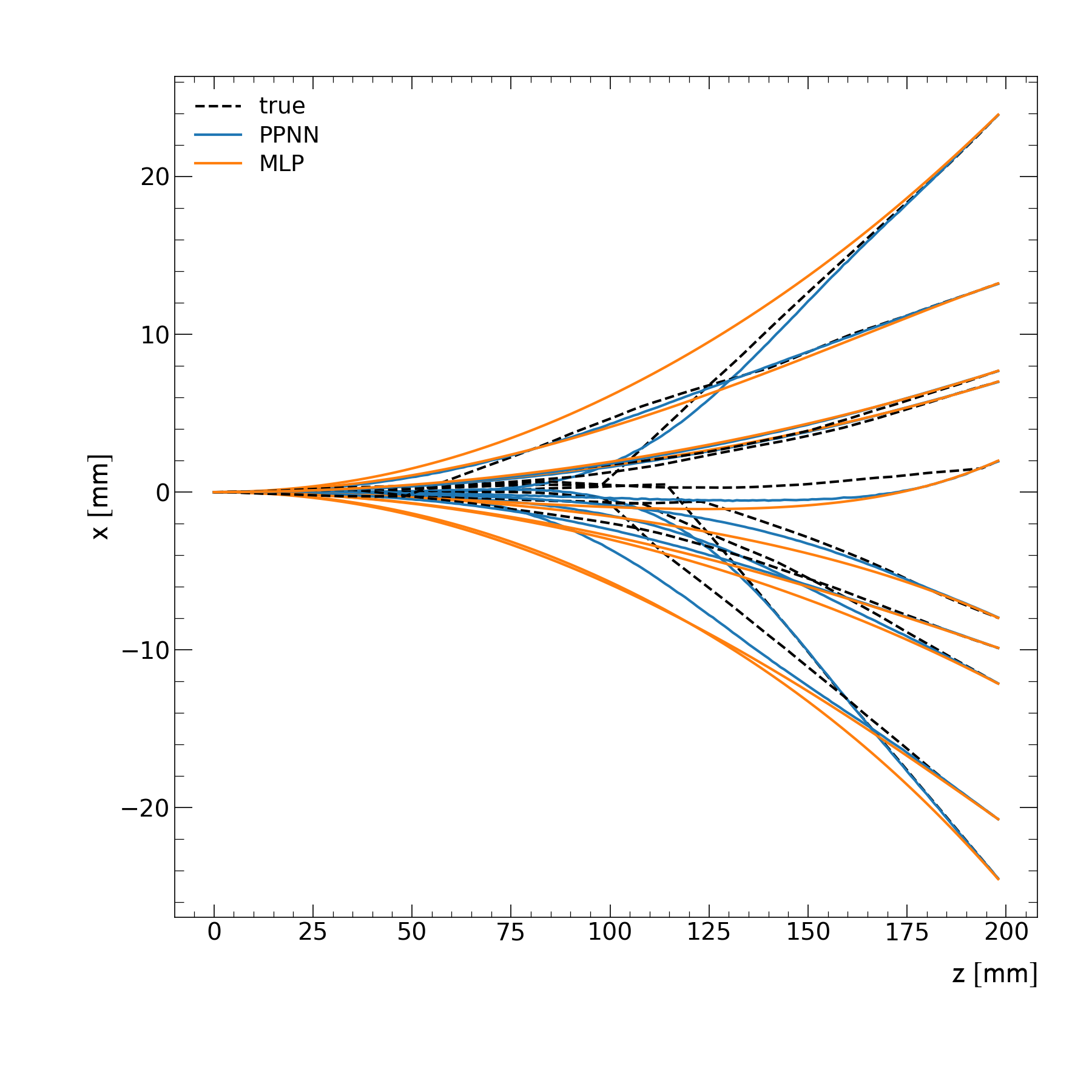}&
\includegraphics[trim={0cm 2.5cm 0cm 0cm},clip,width=.5\linewidth]{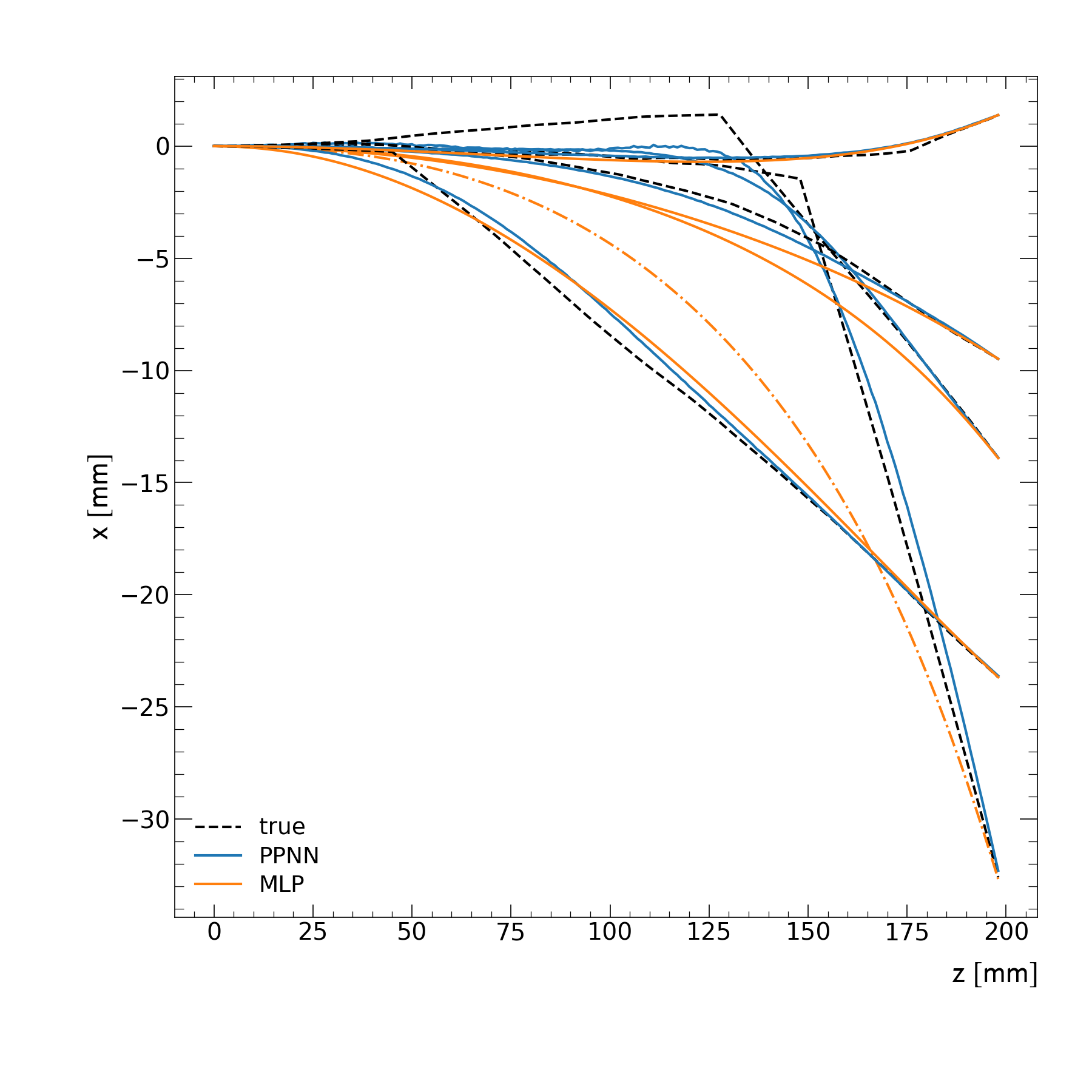}\\
(\textbf{a}) &
(\textbf{b})
\end{tabular}
\caption{Examples of tracks for which the \netname outperform MLP. (a) Tracks are selected at random from inside each of 10 quantiles, using the data of Figure~\ref{fig:delta_RMSE}. (b) Same as (a), but in which tracks are extracted from quartile groups; with the last quartile, which corresponds to tracks with the largest discrepancies between the two methods, divided into two.}
\label{fig:examples}
\end{figure}

Focusing in on only the behaviour when \netname outperforms MLP, let us consider only the set of events on the negative side of histogram. Dividing into 10 quantiles split by $\Delta RMSE$, in Figure~\ref{fig:examples}-(a) we illustrate a selection of randomly chosen tracks, one from each quantile. As expected, for larger deviations from straight paths \netname can better follow the simulated curve in the majority of such cases, growing more notable for larger $\Delta RMSE$. For Figure~\ref{fig:examples}-(b) the same dataset is divided into quartiles, with the last bin, containing tracks with the largest error difference, further divided into two subgroups. As with Figure~\ref{fig:examples}-(a) we chose a random track from each of the five groups. Both figures further support that \netname improved performance is due at-least in part to a better capability to reproduce the particle path in the presence of nuclear interaction, which causes greater changes in the direction of the track.

\begin{figure}[t]
\centering
\begin{tabular}{cc}
\includegraphics[trim={0cm 0.7cm 0cm 0cm},clip,width=.5\linewidth]{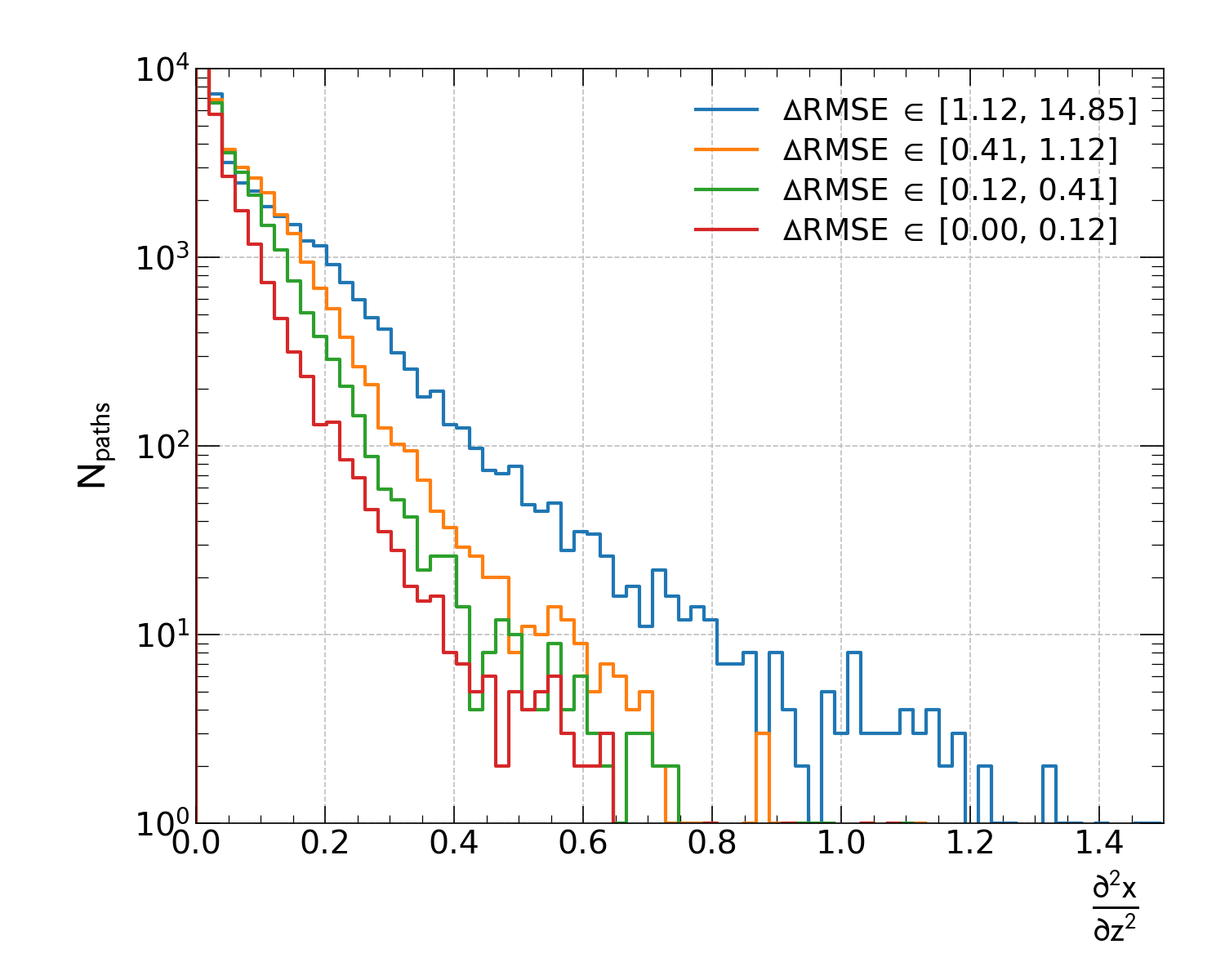}&
\includegraphics[trim={0cm 0.7cm 0cm 0cm},clip,width=.5\linewidth]{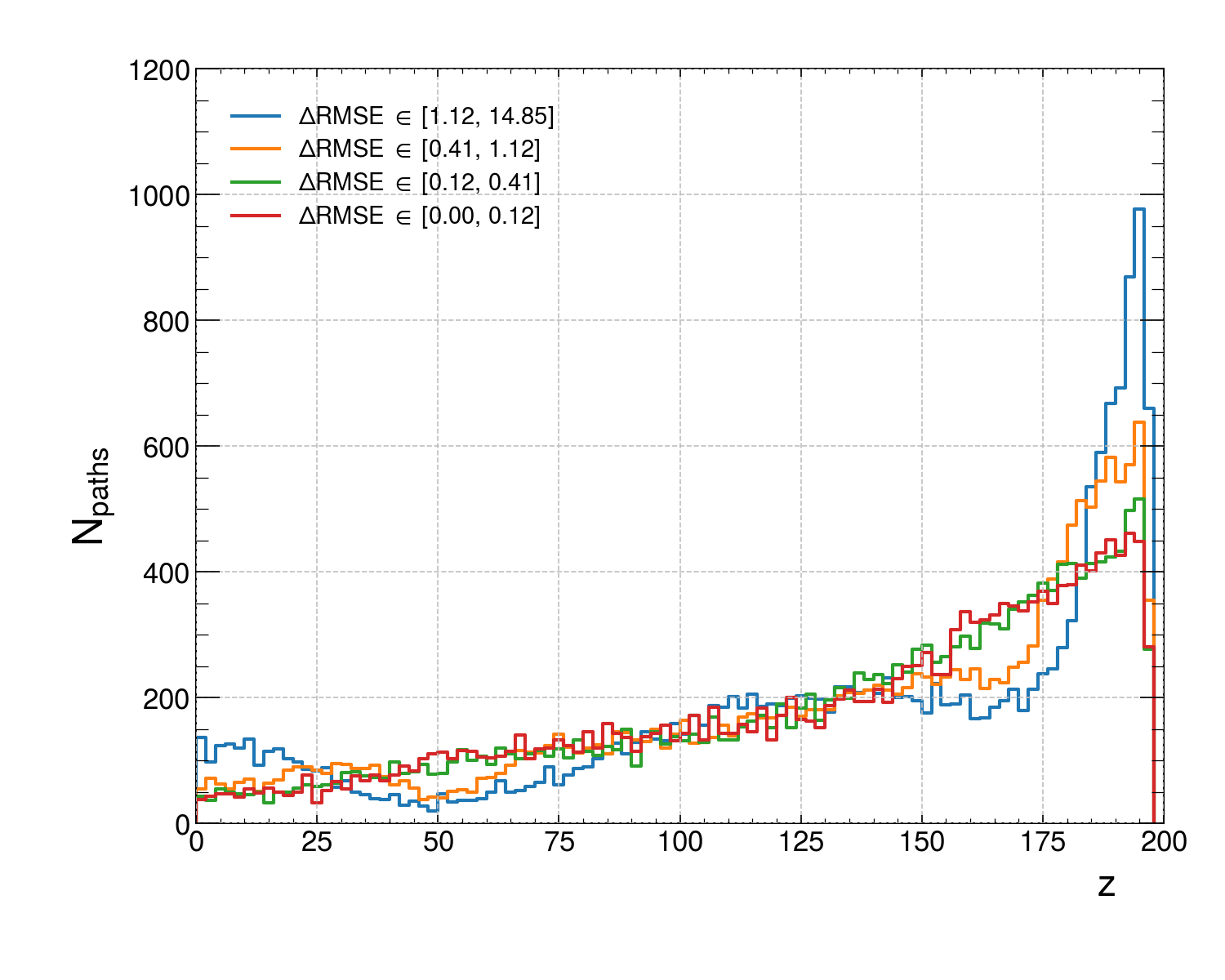}\\
(\textbf{a}) &
(\textbf{b})
\end{tabular}
\caption{(a) Distributions of the second derivative of the tracks in the $x$ direction with respect to the $z$ coordinate. Lines indicate the four quartiles of the distribution of $\Delta RMSE < 0$. (b) Distribution of the position along the $z$ axis for the maximum of the second derivative for each path.} \label{fig:second_der}
\end{figure}

To further analyse this characteristic, Figure~\ref{fig:second_der}-(a) shows the distribution of the second derivative of the $x$ component of the tracks, with respect to the $z$ direction, again for track in which \netname outperforms MLP, broken down into quartiles. Large values of this quantity are connected with significant direction change, such as those observed in Figure~\ref{fig:examples}. The four lines correspond to the four quartiles of the blue histogram in Figure~\ref{fig:delta_RMSE}, as introduced in Figure~\ref{fig:examples}-(b). Where \netname exhibits the better performance, we see that the difference between the tracks reconstructed with \netname and MLP grows with increasing values of $\frac{\partial^2x}{\partial z^2}$: the more a trajectory differs from pure MCS scattering, the more the \netname improves over MLP. 

Figure~\ref{fig:second_der}-(b) shows the distribution of $\max(\frac{\partial^2x}{\partial z^2})$ as a function of $z$. The distribution for the last quartile, corresponding to the largest discrepancies between the two methods, has a notably different behaviour compared to the other three lines. It exhibits significantly more events occurring at small and large $z$ values. An example of these events can be seen in Figure~\ref{fig:examples}-(b) where we have a strong deflection at $z\approx190$ mm. We see that MLP struggles to reproduce this event while the neural network can provide a superior result.

\subsection{Inhomogeneous slab phantom}\label{sec:inhomo}

\begin{figure}[tb]
\centering
\begin{tabular}{cc}
\includegraphics[trim={0cm 2cm 0cm 0cm},clip,width=.5\linewidth]{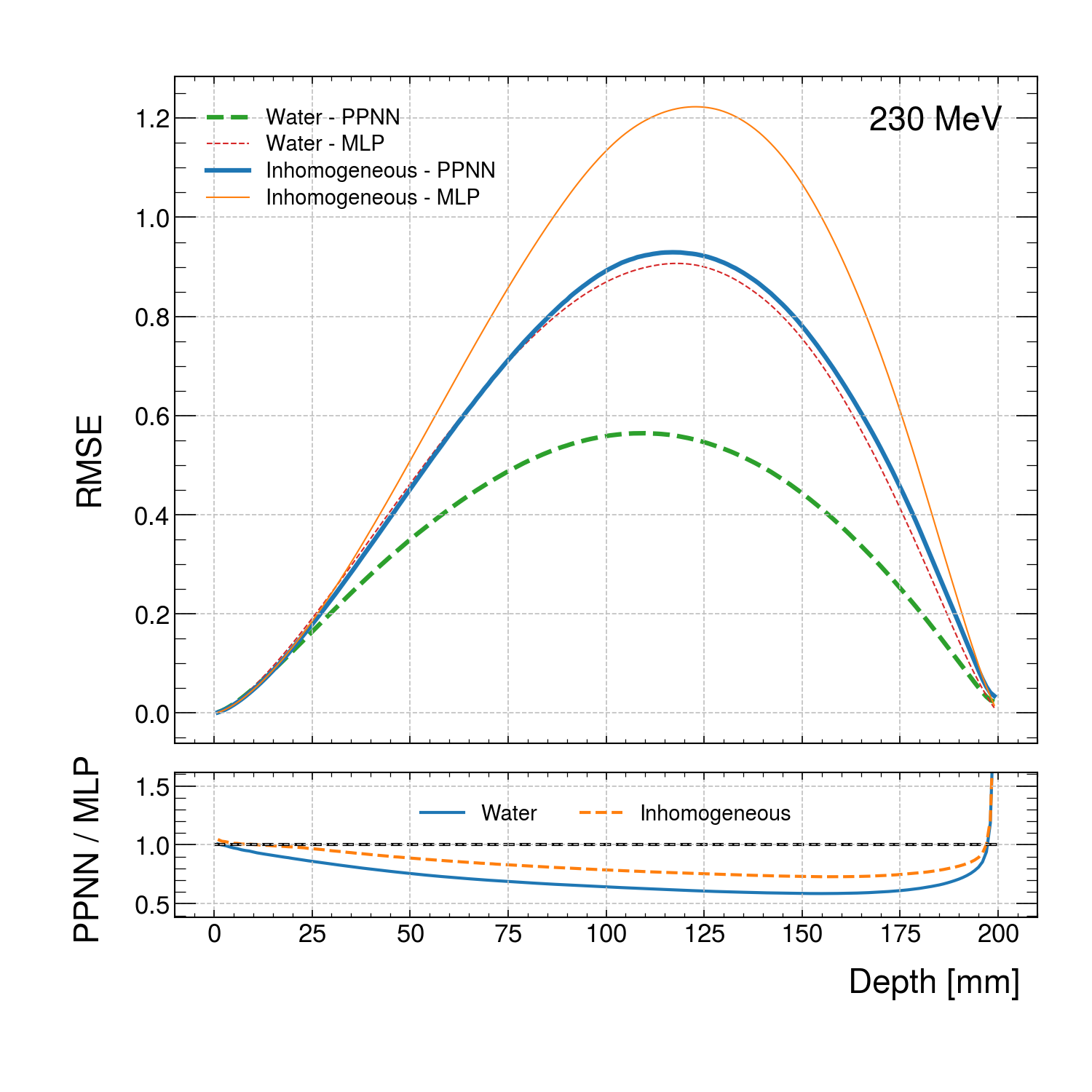}&
\includegraphics[trim={0cm 2cm 0cm 0cm},clip,width=.5\linewidth]{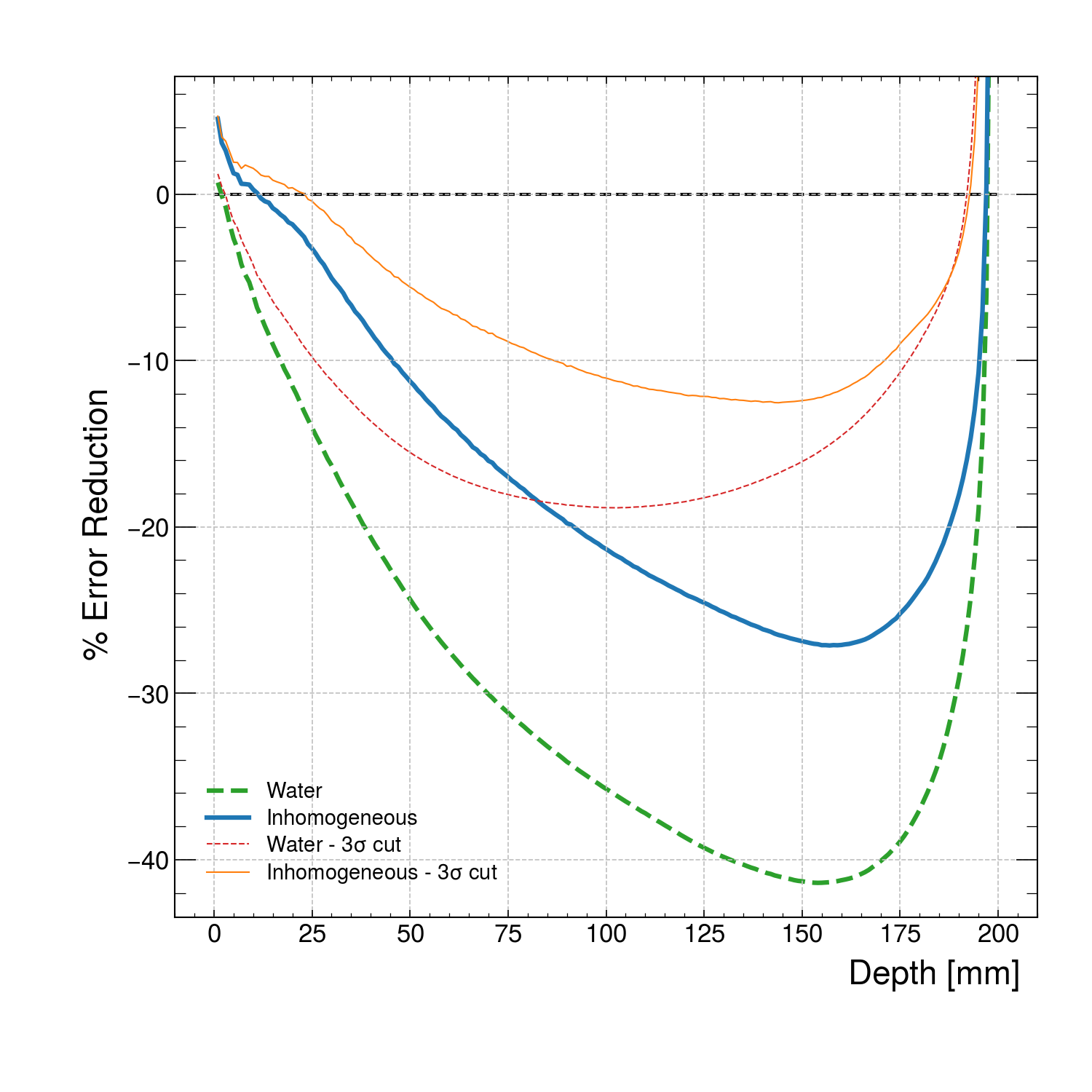}\\
(\textbf{a}) &
(\textbf{b})
\end{tabular}
\caption{(a) Root Mean Squared Error obtained with MLP and \netname on a water and an inhomogeneous slab phantom irradiated at $230$~MeV. (b) Percentage reduction in RMSE with respect to depth by \netname over MLP. All studies were performed under the $QGSP\_BIC$ physics environment.} \label{fig:230mev}
\end{figure}

In this section, we present the results obtained using \netname in the reconstruction of proton trajectory traversing the slab phantom described in \ref{sec:mc_simul} and represented schematically in Figure~\ref{fig:mc_setup}-(c). Due to the inhomogeneous phantom's increased stopping power, a proton energy of $230$~MeV was chosen to ensure a significant fraction of simulated events traversed the full phantom depth. This ensured datasets in excess of $1,400,000$ trajectories ($700,000$ along each direction) for $10^6$ simulated particles. Both \netname and MLP methods were re-trained (re-calibrated for MLP) to the new energy scheme, as described in Sections~\ref{sec:network} and \ref{sec:mlp}. For this purpose, we consider a simulation with $230$~MeV protons through a water phantom analogous to the one used in the $200$~MeV case.

The RMSE error for both phantoms, using either \netname or MLP, is shown in Figure~\ref{fig:230mev}-(a). This compares the water and inhomogeneous systems, without cuts and using the $QGSP\_BIC$ physics environment. For the water phantom both \netname and MLP behave similarly to the corresponding $200$~MeV case. This is an important check that the higher energy implementations of the two methods are functioning correctly.

Focusing on the reconstruction error for the inhomogeneous case, we similarly observe that with \netname the error is consistently reduced. Interestingly the error on the new phantom using \netname is comparable with that obtained with MLP in the pure water simulation. 

The improvement obtained with \netname is more pronounced when examining the percentage reduction of RMSE by \netname over MLP, as shown in Figure~\ref{fig:230mev}-(b). A reduction in the error of the order of $25\%$ can be seen around 150mm, while on average the improvement is in excess of $10\%$ over MLP across a significant portion of the depth. Introducing the familiar $3\sigma$ cuts decreases the error reduction in both the water and inhomogeneous cases, along with the difference in improvement between them.

\subsection{Execution time comparison}\label{sec:times}

For this comparison of the execution time of the two algorithms, the highly optimized version of MLP presented in \citet{mcallister2009efficient} is used, in which $90\%$ of the MLP is precalculated and the number of operation required is minimized. We ported the code in python using the vectorization capabilities of the NumPy (numpy.org) library to parallelize the execution on the number of protons. \netname is written in python using the PyTorch (pytorch.org) framework.

Both codes were executed on the CPU of a Standard NC6 Microsoft Azure machine. Running the two algorithms on all the 1,600,000 trajectories of the test dataset in unique batch combinations and repeating the procedure 10 times we obtain an almost constant execution time of $0.47\pm 0.01$ sec for \netname and $7.11\pm 0.08$ sec for MLP. Within the validity of this test, the \netname method is sixteen times faster than the optimized MLP.

\section{Discussion} \label{sec:discussion}
Although MLP represents a powerful method of estimating proton path in pCT applications, it suffers from different limitations. The approach is designed specifically to account only for effects on the proton path connected with MCS and energy loss. This is reflected by the strategy of discarding protons trajectories with large deviation from straight paths to reduce the error. Moreover, simulation in a realistic scenario of high fluence (hundreds of millions of protons) and small spacing for the MLP (fraction of millimetre) can require more than one hour; time mostly spent reconstructing the proton (paths~\citet{khellaf2020comparison}).

In the interests of alleviating these two problems we propose an alternative method, based on Deep Learning Neural Network,  to estimate the proton trajectory for pCT. The results presented in the previous section suggests that within the \netname approach, these two problems can be relieved to some degree. Figure~\ref{fig:rmse} and Figure~\ref{fig:delta_RMSE} show that using \netname a good approximation of the path can be obtained for a much larger number of protons than using MLP. This is important because in principle fewer protons are needed to reach the same reconstruction quality, lowering both the dose and  the computation time. Consolidating this claim is one of the aims of our future developments.

The ability of the network to reconstruct tracks outside the validity of the MLP approach is intrinsically tied to the nature of deep learning. Neural networks learn "blindly" from examples; parsing though the training dataset, by means of the back-propagation procedure for the minimisation of the loss function, the network adapts its weights to the characteristics of the events it experiences, including those that show large $\Delta\theta$ and/or $\Delta x$. While such underlying processes maybe challenging to formulate into mathematical models, there are sufficient patterns for the network to refine its prediction processes. Without an assumed structure to reproduce, it is not bound to solely replicating the form of a given physical model. A tentative explanation of what the network learns may be inferred from Figure~\ref{fig:second_der} and the analysis of the second derivative of $x$ w.r.t. $z$. The network displays significant improvement over MLP where the second derivative is large, especially near the end of the trajectories.

The study of inhomogeneous systems is only started here, and it certainly warrants a much more in-depth investigation into more realistic configurations of the phantom. The phantom considered is certainly extreme; large volumes of a high-density material such as those in the slab phantom will rarely be encountered in clinical practice, and in this sense we do not expect the gain to be so large in a realistic situation. Nevertheless, it is encouraging that notably better results are obtained with \netname with respect to MLP, with a reductions of the RMSE of the order of $20\%$. This is a more significant improvement compared to the work presented in \cite{brooke2020inhomogeneous} with a similar phantom, where the maximum enhancement is about $5\%$ for simulation with the same beam energy.

Regarding execution speed, it is true that the time spent for reconstruction is only one of the various aspects for evaluating a pCT system for clinical routine. Moreover, our work is relevant only in the context of reconstruction methods based on the evaluation of the proton path. Nevertheless, because these methods are seen as the most promising for applicability in the clinical context and the MLP execution speed is by order of magnitudes the slowest part of the algorithm (\citet{khellaf2020comparison}), the substantial improvement shown by \netname compared with the optimized MLP can be regarded as an important feature.

\section{Conclusions}\label{sec:conclusion}

MLP is the principal method adopted in pCT for the reconstruction of single proton paths through the body. In this paper we have demonstrated that using Deep Learning Neural Network it is possible to recreate the same performance of MLP in the regime in which MLP is applicable and achieve a better performance outside its region of validity. Using \netname would also permit discarding fewer protons in the pCT procedure. Moreover, an execution time test of the two algorithms indicates that \netname can be substantially faster in performing the reconstruction. In the future we plan to move forward in the development of the method towards a full reconstruction procedure applicable to more realistic phantoms. 

\section{Acknowledgments}\label{sec:acknowledgments}

We would like to acknowledge Simon Rit for the useful discussion and clarification of the MLP method in the early phase of the experiments and development of the \netname method.
\\

\bibliographystyle{dcu}
\bibliography{pct_path_nn}

\end{document}